\documentstyle[twocolumn,prc,aps]{revtex}
\input epsf
\begin{document}
\twocolumn[\hsize\textwidth\columnwidth\hsize\csname @twocolumnfalse\endcsname 
\title{Baryon Fluctuations in High Energy Nuclear Collisions} 
\author{Sean Gavin and Claude Pruneau}
\address{
Department of Physics and Astronomy, Wayne State University, 
Detroit, MI, 48202}
\date{\today} 
\maketitle
\begin{abstract}
  We propose that dramatic changes in the variances and covariance of
  protons and antiprotons can result if baryons approach chemical
  equilibrium in nuclear collisions at RHIC.  To explore how
  equilibration alters these fluctuations, we formulate both
  equilibrium and nonequilibrium hadrochemical descriptions of baryon
  evolution.  Contributions to fluctuations from impact parameter
  averaging and finite acceptance in nuclear collisions are
  numerically simulated.  \vspace{0.1in}

\end{abstract}
]

\begin{narrowtext}


Event-by-event fluctuations of the particle yields in relativistic
nuclear collisions can be sensitive to the degree of chemical
equilibration \cite{Mrow} and to critical fluctuations due to the QCD
phase transition \cite{Krishna,BaymHeiselberg}. We study for the first
time the degree to which chemical equilibration can affect the
fluctuations of baryons at RHIC and LHC. Baryons and antibaryons are
likely produced in such abundance that event-by-event yields can be
measured \cite{gp1}.  We propose that chemical equilibration can
appreciably change the variances and covariance of protons and
antiprotons compared to thermal and participant-nucleon model
expectations.

Fluctuations of the baryon-antibaryon system can vary depending on the
degree to which the chemical reactions $N\overline{N}
\leftrightarrow$~mesons reach equilibrium \cite{gp1}.  In the absence
of chemical equilibrium, the number of antibaryons and baryons are
independently conserved. Fluctuations of the particle number are then
Poissonian,
\begin{equation} {\cal V}\equiv \sum_i(N_i - \langle N\rangle)^2
\approx  N, \,\,\,\,\,\,\,\,\,\,
\,\,\,\,\,\, {\rm thermal}
\label{eq:eq1}\end{equation} 
where $N_i$ is the baryon number in the $i^{\rm th}$ event and $N
\equiv \langle N\rangle$ is the event average.  HIJING and similar
participant nucleon models exhibit essentially the same behavior for
nucleus-nucleus collisions \cite{BaymHeiselberg}. In contrast,
chemical equilibrium relates fluctuations of the baryons to those of
the antibaryons, reducing the relative variance compared to
(\ref{eq:eq1}).  We find \cite{gp1}:
\begin{equation}
{\cal V}\approx
 N^2(N+\overline{N})^{-1},\,\,\,\,\,\,\,\,\,\, 
 {\rm chemical}
\label{eq:eq2}\end{equation}
where $\overline{N}$ is the average number of antibaryons.
We recover (\ref{eq:eq1}) in the limit $N \gg \overline{N}$, since
baryon conservation fixes the number of particles in that limit.  

If chemical equilibrium is achieved in RHIC collisions, the variance
(\ref{eq:eq2}) can fall to {\em half} the thermal and
participant-nucleon model level, because the numbers of baryons and
antibaryons are expected to be comparable at midrapidity
\cite{vance}. We propose that a further signal of equilibration can be
obtained from the baryon-antibaryon covariance,
\begin{equation}
{\cal C} =
\sum_i (\overline{N}_i - \langle \overline{N}\rangle)
(N_i - \langle N\rangle).
\label{eq:i3}\end{equation}
In chemical equilibrium, we find that the covariance can be negative,
\begin{equation}
{\cal C} \approx
-N\overline{N}(N+\overline{N})^{-1}.
\,\,\,\,\,\,\,\,\,\, 
 {\rm chemical}
\label{eq:i4}\end{equation}
On the other hand, this quantity essentially vanishes for collisions
of large nuclei in thermal equilibrium, or when $N \gg \overline{N}$.

The anti-correlation (\ref{eq:i4}) seems somewhat surprising --
positive correlations between baryons and antibaryons are seen in
$e^+e^-$ experiments \cite{tasso} and incorporated, e.g., in string
fragmentation models \cite{thor}. Such correlations are required by
baryon number conservation for those comparatively small systems. In
contrast, the anti-correlation (\ref{eq:i4}) and the reduction of the
variance (\ref{eq:eq2}) result from fluctuations of the net baryon
number itself. In a subvolume of a large system, fluctuations can
increase the net baryon number $N_B$ if either a particle enters or
an antiparticle exits the subvolume. Correspondingly, fluctuations of $N_B$
simultaneously increase $N$ and reduce $\overline{N}$, leading to an
anti-correlation.  This effect is only possible if chemical reactions
render the individual numbers $N$ and $\overline{N}$ indefinite. 

To determine the degree of chemical equilibration from fluctuations in
experiments, one must account for the following additional sources of
fluctuations, all of which tend to increase $\cal V$ and drive $\cal
C$ toward positive values. Near local equilibrium, thermal and volume
fluctuations produce particle number fluctuations.  In experiments,
further fluctuations result from centrality selection and detector
acceptance effects \cite{gp1,BaymHeiselberg}.  To illustrate how
volume and thermal fluctuations can change the variance of particle
yields, we begin by extending the thermodynamic formulation of
ref.~\cite{gp1}.  We then apply the Monte Carlo event generator of
\cite{gp1} to compute the experimental contribution.

Our results imply that novel correlations can be observed if chemical
equilibrium is obtained. However, it is not obvious that baryons will
achieve chemical equilibrium or that equilibrium correlations will
survive chemical and thermal freezeout. We develop a simple
hadrochemical model to study the approach toward chemical equilibrium.
We estimate the corrections to (\ref{eq:eq1}) and (\ref{eq:eq2}) that
arise if the system is only partially equilibrated. We then briefly
discuss how elastic scattering following chemical freezeout can alter
$\cal C$ and $\cal V$

To illustrate how fluctuations alter antibaryon production in local
thermal and chemical equilibrium, we employ the idealized but standard
Bjorken hydrodynamic framework. We suppose that the energy and baryon
number are deposited near midrapidity by pre-equilibrium processes
that locally establish an initial temperature $T$, entropy density $s$
and net baryon density $\rho_B$.  These quantities are further assumed
to vary only with proper time $\tau$. Entropy and baryon number
conservation then imply that the rapidity densities for entropy,
$S = d{\cal S}/dy$, and net baryon number $N_B = d{\cal N}_B/dy$ are
constant.  The rapidity density of all hadrons is also nearly
constant, because $N_{\rm tot} \propto S$ for a system dominated by
light hadron species with masses $\ll T$. Mean rapidity densities of
individual species, such as antibaryons $\overline N$ and baryons $N$
vary with $\tau$. For this Bjorken scenario, it is useful to define an
effective volume $V \approx {\cal A} \tau$ that increases from a
formation time $\tau_0$ to freezeout at $\tau_F$, such that $V = S/s
\propto N_{\rm tot}/n_{\rm tot}$.  The transverse area $\cal A$ is
initially determined by the overlap of the colliding nuclei.

In local thermal and chemical equilibrium, the variances and
covariances of the state variables specify the fluctuations of all
bulk thermodynamic quantities. For our collision scenario, the
appropriate state variables are $T$, $V$ and $N_B$.  The fluctuations
of a quantity $X$ are characterized by the variance $\sigma_{X}^2 =
\langle\Delta X^2\rangle$ for $\Delta X \equiv X - \langle X\rangle$.
If we take the total number of hadrons $N_{\rm tot}$ to be
proportional to the entropy, then
\begin{equation}
\sigma_V^2/V^2
\approx
\langle\Delta N_{\rm tot}^2\rangle/N_{\rm tot}^2
\approx N_{\rm tot}^{-1}.
\label{eq:eq4a}\end{equation}
As in \cite{gp1}, we take the fluctuations of $T$ and $N_B$ to 
satisfy:
\begin{equation}
\sigma_T^2 = T^2C_v^{-1}, 
\,\,\,\,\, \,\,\,\,\,    
\sigma_{B}^2 = T\partial N_B/\partial \mu_B 
\label{eq:eq4b}\end{equation}
for a system with a heat capacity $C_v$; see e.g. ref.~\cite{LL1}. For
an ideal hadron gas,
\begin{equation}
\sigma_T^2 \approx 
T^2\sigma_V^2/12V^2,
\,\,\,\,\, \,\,\,\,\,    
\sigma_{B}^2 = N + \overline{N},
\label{eq:eq4c}\end{equation}
neglecting small corrections from Fermi and Bose statistics.  To
completely specify our ensemble, we assume that fluctuations of these
state variables are statistically independent, so that
$\langle\Delta V\Delta T\rangle = \langle \Delta N_B\Delta T\rangle =
\langle\Delta N_B \Delta V\rangle \equiv 0.$
Observe that while this formulation gives the following the flavor of
a thermodynamic analysis, ours is not a static, global equilibrium
scenario. 

Volume and thermal fluctuations cause the number of baryons and
antibaryons to fluctuate \cite{gp1}. In the absence of chemical
equilibrium, we write:
\begin{equation}
\Delta N = N\Delta V/V + \Delta N_{{}_{VT}},
\label{eq:eq5}\end{equation}
with a similar equation for antibaryons. The contribution $\Delta
N_{{}_{VT}}$ for constant $T$ and $V$ is (\ref{eq:eq1}).  Thermal
fluctuations do not change the conserved particle numbers, while
volume fluctuations add to (\ref{eq:eq1}). The variance is
\begin{equation}
{\cal V}_{\rm th} \equiv \sigma_{N}^2
=
N^2\sigma_V^2/V^2
+ N \approx N(1+N/N_{\rm tot}).
\label{eq:eq6}\end{equation}
We also compute the baryon-antibaryon covariance:
\begin{equation}
{\cal C}_{\rm th} 
\equiv
\langle \Delta N \Delta\overline{N}\rangle
= N\overline{N}\sigma_V^2/V^2 
\approx N\overline{N}/N_{\rm tot}.
\label{eq:eq9}\end{equation}
This result implies that given $N$ particles, the probability of
finding an antiparticle is $\overline{N}/N_{\rm tot}$.  The correction
to (\ref{eq:eq1}) and (\ref{eq:i4}) from volume fluctuations are of
order 4\%, since $N/N_{\rm tot}\sim 0.04$ for central Au+Au at RHIC.

In chemical equilibrium, the number of particles and antiparticles
obey:
\begin{equation}
\Delta N = 
{{N\Delta V}\over{V}} +
\left({{\partial N}\over {\partial T}}\right)_{{}_{N_B}}
\!\!\!\!{\Delta T}
+ 
\left({{\partial N}\over {\partial N_B}}\right)_{{}_{T}}
\Delta N_B,
\label{eq:eq10}\end{equation}
and
\begin{equation}
\Delta \overline{N} = 
{{{\overline N}\Delta V}\over{V}} + 
\left({{\partial {\overline N}}\over {\partial T}}\right)_{{}_{N_B}}
\!\!\!\!\Delta T
+ 
\left({{\partial {\overline N}}\over {\partial N_B}}\right)_{{}_{T}}
\Delta N_B.
\label{eq:eq11}\end{equation}
where $(\partial N/\partial T)_{N_B}=(\partial \overline{N}/\partial
T)_{N_B}$ and $(\partial N/\partial N_B)_T-(\partial
\overline{N}/\partial N_B)_T = 1$.  In chemical equilibrium, thermal
fluctuations can change the baryon population by making pairs, so that
\begin{equation}
\left({{\partial N}\over {\partial T}}\right)_{{}_{N_B}}
= {{\partial N}\over {\partial T}}
- \left({{\partial N}\over {\partial \mu_B}}\right)
{{\partial N_B/\partial T}\over {\partial N_B/\mu_B}}.
\label{eq:eq12}\end{equation}
For an ideal gas,
\begin{equation}
\Delta N = 
N{{\Delta V}\over{V}}  
+{{2\epsilon N\overline{N}}\over {N +\overline{N}}}{{\Delta T}\over T}
+ {{N\Delta N_B}\over {N+\overline{N}}}
\label{eq:eq13}\end{equation}
and
\begin{equation}
\Delta \overline{N} = 
\overline{N}{{\Delta V}\over{V}}  
+{{2\epsilon N\overline{N}}\over {N +\overline{N}}}{{\Delta T}\over T}
- {{\overline{N}\Delta N_B}\over {N+\overline{N}}}.
\label{eq:eq13a}\end{equation}
where $\epsilon = E_{N}/NT \approx m/T + 3/2$ is
the energy per antibaryon per unit temperature. 
We then obtain:
\begin{equation}
{\cal V}_{\rm ch} =
N^2{{\sigma_V^2}\over{V^2}} + 
4\epsilon^2\left({{N\overline{N}}\over {N+\overline{N}}}\right)^2
{{\sigma_T^2}\over {T^2}}
+ {{N^2}\over {N+\overline{N}}}.
\label{eq:eq14}\end{equation}
For $N \gg \overline{N}$ we obtain (\ref{eq:eq6}). 
The covariance is:
\begin{equation}
{\cal C}_{\rm ch} =
N\overline{N}{{\sigma_V^2}\over{V^2}}
+
4\epsilon^2\left({{N\overline{N}}\over{N+\overline{N}}}\right)^2
{{\sigma_T^2}\over{T^2}}
-
{{N\overline{N}}\over{N+\overline{N}}}
\label{eq:eq18}\end{equation}
The first terms in (\ref{eq:eq14}, \ref{eq:eq18}) are the volume
contributions encountered earlier. These terms are unchanged by
chemical equilibrium.  The second term accounts for the pairwise
production of baryons and antibaryons by thermal fluctuations. The
third term describes the net baryon number fluctuations discussed
earlier, see eq.~(\ref{eq:i4}).

At RHIC, $N\approx \overline{N}$ implies that both the baryon variance
and the baryon-antibaryon covariance markedly differ from the
nonequilibrium results. The sum of the volume and thermal terms is
roughly $1 + \epsilon^2/12\approx 5.5$ for chemical freezeout at $T =
160$~MeV and $m = 938$~MeV. The contribution from the first two terms
to either ${\cal V}_{\rm ch}/N^2$ or ${\cal C}_{\rm ch}/N\overline{N}$
is then $\sim 0.5\%$ for $N_{\rm tot} \sim 10^3$ as expected in Au+Au
collisions.  Baryon density fluctuations contribute $\sim +1.3\%$ to
the variance and $-1.3\%$ to the covariance for $\overline{N} \approx
N\approx 40$, yielding the totals ${\cal V}_{\rm ch}/N^2\sim 1.8\%$
and ${\cal C}_{\rm ch}/N\overline{N}\sim -0.8\%$.  Observe that the
chemical nonequilibrium results (\ref{eq:eq6}, \ref{eq:eq9}) give
values $\sim 2.5\%$ and $+0.1\%$ for the relative variance and
covariance; we expect participant nucleon models to yield similar
values. We will see that that these small percentage-differences
amount to quite large changes in the magnitudes of these quantities.

We comment that particle ratios, such as $N/N_{\rm tot}$, are often
measured in place of absolute yields. Variances of ratios do not
contain the contribution from volume fluctuations found e.g., in
(\ref{eq:eq6}) and (\ref{eq:eq14}). However, experimenters must
carefully construct ratios from $N$ and $N_{\rm tot}$ from data
measured over the full kinematic ranges of these particles in order to
fully cancel the volume fluctuations and to suppress new fluctuations
due to systematic shifts in the spectra.  The distinction between
ratios and yields is minor, however, since we find that volume
fluctuations are the smallest of the contributions treated here.

To account for additional sources of fluctuations in heavy ion
collisions, we incorporate these general results into the Monte Carlo
event generator for correlated signals developed in \cite{gp1}.
There, we assume that the rapidity densities $N$, $\overline N$ and
$N_{\rm tot}$ can be described by a thermal ensemble at each impact
parameter $b$, with fluctuations occurring about the well defined mean
values obtained below. We then generate events, choosing the values of
$\overline{N}_i$ and $N_{i}$ for the $i^{\rm th}$ event using
(\ref{eq:eq5}), (\ref{eq:eq10}, \ref{eq:eq11}), or (\ref{eq:fluct3}),
depending on whether the system is in thermal equilibrium, chemical
equilibrium, or in between.  To simulate the centrality dependence of
ion collisions, we distribute events in $b$ according to a wounded
nucleon model with realistic nuclear density profiles.

We compute the mean rapidity densities of protons and antiprotons for
Au+Au at RHIC using the wounded nucleon model relations,
\begin{equation}
N \approx n{\cal N}(b)/{\cal N}(0),
\,\,\,\,\,\,\,\,\,\,\,
\overline{N} \approx \overline{n}{\cal N}(b)/{\cal N}(0),
\label{eq:pbar}
\end{equation}
where ${\cal N}(b)$ is the number of participant nucleons. We take
values of the rapidity density at zero impact parameter $n \approx
\overline{n} \approx 40$ for both baryons and antibaryons. Such values
are consistent with the range of event generator predictions: HIJING,
HIJING/$B{\overline B}$ (its successor) and RQMD report rapidity
densities of 60, 20 and 20 respectively. [Note that we will use
(\ref{eq:pbar}) to provide initial conditions when we consider partial
equilibrium effects, see eq.~(\ref{eq:ans2}).]  We then compute the
average charged particle multiplicity for each event assuming a
Gaussian distribution with an average value $N_{\rm tot}(b) = N_{\rm
  tot}(0){\cal N}(b)/{\cal N}(0)$ and a standard deviation
$\sigma_{\rm tot} = \sqrt{N_{\rm tot}}$ consistent with thermal
equilibrium. The scale $N_{\rm tot}(0) = 2100$ is determined by the
initial production regardless of event class, in accord with entropy
and energy conservation; the particular value is taken from a HIJING
simulation in the STAR acceptance.

The negative covariance that signals chemical equilibrium is not
destroyed by volume, thermal, finite-acceptance, or impact-parameter
fluctuations.  In fig.~1, we show the proton variance and
proton-antiproton covariance computed from $10^6$ events in the STAR
acceptance assuming that chemical equilibrium is complete. Such an
event sample can be accumulated in roughly twelve days of STAR running
and can provide a statistically significant determination of $\cal{C}$
and $\cal{V}$ as functions of the multiplicity. Chemical equilibrium
fluctuations are computed using (\ref{eq:eq10}, \ref{eq:eq11}).  Our
results shown by the solid curves are strikingly different from
thermal equilibrium expectations -- the long-dashed curves -- obtained
from eq.~(\ref{eq:eq5}) at the same density.  Incorporated in our
simulation is an important non-thermal source of fluctuations that
results from impact parameter averaging, as discussed in \cite{gp1}.
These ``background'' fluctuations, shown as the short-dashed curves,
result from the binning of the variances and covariance as functions
of multiplicity.
\begin{figure} 
\vskip -0.35in
\epsfxsize=4.5in
\centerline{\epsffile{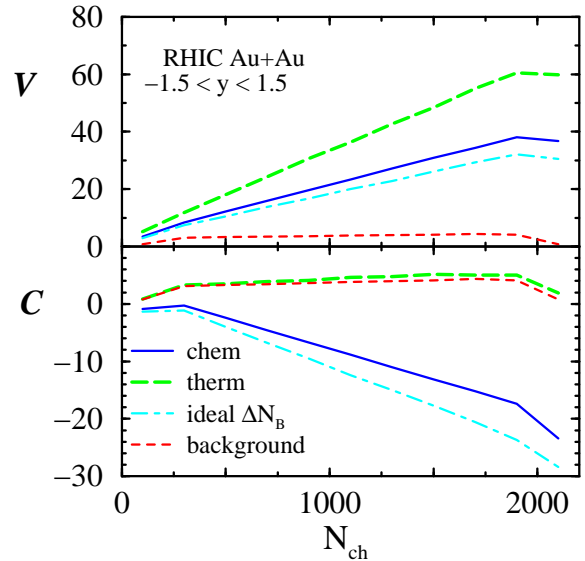}}
\vskip -0.25in 
\caption[]{
  Proton variance (top) and proton-antiproton covariance (bottom) for
  a chemical equilibrium state in RHIC Au+Au. The signature of
  baryon-density fluctuations -- the negative correlation -- remains
  striking in the covariance, despite competition from thermal,
  volume, and impact-parameter fluctuations.}
\end{figure}

Having found that chemical equilibrium for baryons can have observable
consequences, we now ask whether these particles approach chemical
equilibrium in the first place. To be concrete, we suppose that
baryons, antibaryons and other hadrons form at a proper time $\tau_0$
from the hadronization of a quark gluon plasma. These baryons can then
scatter and annihilate, while reactions such as $\rho\omega
\rightarrow N\overline{N}$ produce new baryons.  Chemical equilibrium
is obtained when the annihilation rate precisely matches the creation
rate. 

To describe the approach to equilibrium, we extend the kinetic model
of \cite{ggpv} to include meson-meson production in the following
schematic fashion.  We take the antibaryon and baryon density to
satisfy the approximate kinetic equation:
\begin{equation}
\left({{d}\over{d\tau}}+{{1}\over \tau}\right)n
= -\langle \sigma_a v_{\rm rel}\rangle\left(n {\overline n} - n_{\rm eq} 
{\overline n}_{\rm eq}\right)
\label{eq:k1}\end{equation}
together with baryon number conservation. Again we assume a
Bjorken-like expansion. Detailed balance for $N\overline{N}
\rightleftharpoons$~mesons fixes the product of the chemical
equilibrium densities $n_{\rm eq} {\overline n}_{\rm eq}$ in terms of
the meson densities. Note that this relaxation-time approximation
strictly holds only when the initial baryon, antibaryon and meson
densities are sufficiently close to local chemical equilibrium values.

In terms of the rapidity densities ${\overline N}$ and $N$, we write
\cite{ggpv}:
\begin{equation}
\tau{{dN}/{d\tau}} \approx
-\gamma~\{N {\overline N} - N_{\rm eq} {\overline N}_{\rm eq}\}.
\label{eq:k3}\end{equation}
Baryon conservation implies $N - \overline{N} = N_B$.  The coefficient
$\gamma = \langle \sigma_a v_{\rm rel}\rangle {\cal A}^{-1}$
depends on the collision frequency.  In \cite{ggpv}, we   
%
%
used the measured energy-dependent annihilation cross section to
compute $\langle \sigma_a v_{\rm rel}\rangle\approx $~4.4~fm$^2$. We
take $\overline{N}_{\rm eq}$ to be a $\tau$-independent parameter to
be varied over a range of plausible values. This assumption is reasonable
for $N$, $\overline{N}\ll N_{\rm tot}$, since the change in the meson 
populations due to baryon annihilation is then negligible.

\begin{figure} 
\vskip -0.35in
\epsfxsize=4.0in
\centerline{\epsffile{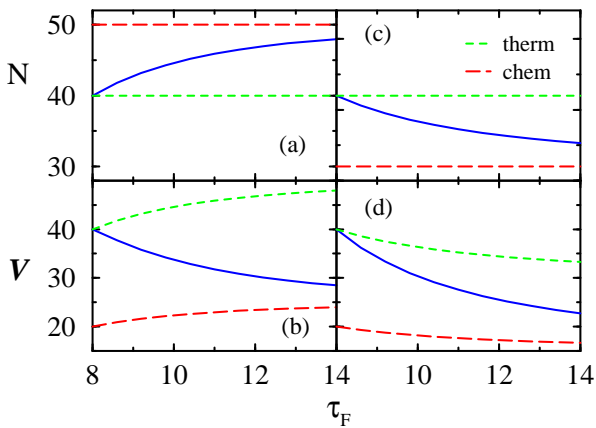}}
\vskip -0.30in 
\caption[]{
  The rapid approach to chemical equilibrium of the proton rapidity
  density (top) and variance (bottom) as functions of the freezeout
  time $\tau_F$ for $N_{\rm eq} > N_0$, (a) and (b), and for $N_{\rm
    eq} < N_0$.  Hadron formation occurs at $\tau_0 = 8$~fm.}
\end{figure}
To solve (\ref{eq:k3}), observe that $N(\tau)$ will increase toward
$N_{\rm eq}$ depending on whether the right hand side of (\ref{eq:k3})
is positive or negative. Growing solutions are obtained if the initial
rapidity density $N_0$ satisfies $N_0 < N_{\rm eq}$, i.e., if the
ratio
\begin{equation}
\theta = {{N_{\rm eq} - N_0}\over
          {N_{\rm eq} + N_0 - N_B}} 
\label{eq:ans1}\end{equation}
is greater than zero. The solution is then:
\begin{equation}
N = {{N_B}\over{2}} +
\left(N_{\rm eq} - {{N_B}\over{2}} \right)
{{1 - \theta S}\over{1 + \theta S}} 
\label{eq:ans2}\end{equation}
where 
\begin{equation}
S =  (\tau_0/\tau_F)^{\gamma(2N_{\rm eq}-N_B)},
\label{eq:ans3}\end{equation}
with chemical freezeout occurring at $\tau_F$.  If $\theta \ll 1$
then annihilation reduces the antiproton-to-proton ratio as in
\cite{ggpv}. The evolution of $N$ toward chemical equilibrium is shown
in figs.~2a and 2c for $N_{\rm eq} > N_0$ and $N_{\rm eq} < N_0$,
respectively.

To estimate fluctuations in the nonequilbrium state, we write the following
ansatz:
\begin{equation}
{\cal V} =
S{\cal V}_{\rm th}
+
(1-S){\cal V}_{\rm ch},
\label{eq:fluct3}\end{equation}
where ${\cal V}_{\rm th}$ and ${\cal V}_{\rm ch}$ respectively satisfy
(\ref{eq:eq6}) and (\ref{eq:eq14}).  To motivate this ansatz, we
consider the variance as a function of $\tau$ and linearize for $N$
near $N_{\rm ch}$, as is standard \cite{LL}.  We find ${\cal V}\sim
{\cal V}_{\rm ch} + 2 N_{\rm eq}(N(\tau)-N_{\rm eq})$. Differentiating
and using (\ref{eq:k3}), we see that $\tau d{\cal V}/d\tau \approx 2\kappa
{\cal V}$, where $\kappa \sim d(\log N) /d(\log \tau)\sim -\gamma$.
The ansatz (\ref{eq:fluct3}) has the correct long- and short-scale
time dependence to linear order, satisfies the appropriate boundary
conditions for $\tau$ near zero and infinity and has plausible
behavior for intermediate $\tau$. The ansatz
\begin{equation}
{\cal C} =
S{\cal C}_{\rm th}
+
(1-S){\cal C}_{\rm ch},
\label{eq:fluct4}\end{equation}
is similarly plausible.

\begin{figure} 
\vskip -0.35in
\epsfxsize=4.5in
\centerline{\epsffile{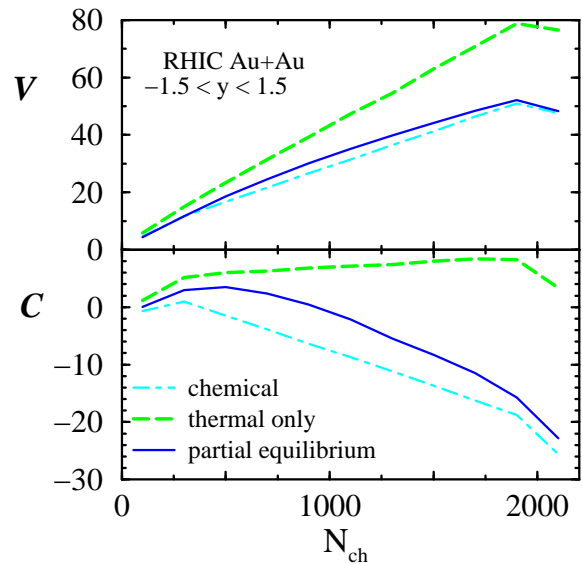}}
\vskip -0.25in  
\caption[]{
  Proton variance (top) and covariance (bottom) for a partial chemical
  equilibrium state in RHIC Au+Au compared to complete chemical
  (dot-dashed) and thermal equilibrium (long dashed) expectations for
  the same $N$ and $\overline N$. }
\end{figure}
To appreciate the speed with which equilibration takes place, we show
the calculated approach to chemical equilibrium of $N$ and $\cal V$ in
fig.~2. We assume that hadronization proceeds from a thermalized quark
gluon plasma and, correspondingly, take $\tau_0 = 8$~fm. The initial
values $N_0 \approx \overline{N}_0$ are taken from (\ref{eq:pbar}) for
$b=0$. The chemical freezeout time is varied up to a value
$\tau_F \approx 14$~fm, a plausible value for central Au+Au at RHIC.
For this figure we take ad hoc values for $N_{\rm eq} \approx
\overline{N}_{\rm eq}$ to illustrate the approach to equilibrium from
above and below.  Equilibration is rapid because the annihilation
cross section and baryon rapidity densities are sufficiently large
that the ratio of the collision time to the expansion time scale
satisfies $\tau_{\rm coll}/ \tau_{\rm exp}\sim (2\gamma N_{\rm
  eq})^{-1} < 1$.

The variance and covariance computed for the STAR acceptance using
$10^6$ events simulated according to our partial equilibrium solution
are shown in fig.~3.  Rapidity densities are computed using
(\ref{eq:ans1} - \ref{eq:ans3}) for initial rapidity densities from
(\ref{eq:pbar}). In these calculations, we take the ad hoc value
$N_{\rm eq} = 50$ for central collisions and assume $N_{\rm eq}$
scales with the number of participants.  The chemical freezeout time
is computed for each impact parameter using $\tau_F = R/v_s +
\tau_p$, where $R = ({\cal A}/\pi)^{1/2}$ is the geometric transverse
radius, $v_s = 1/\sqrt{3}$ is the sound speed and $\tau_p \approx
1$~fm is a limiting value for peripheral collisions set by the proton
size. Using these values, fluctuations are simulated in accord with
(\ref{eq:fluct3}, \ref{eq:fluct4}).  We again take the formation time
for hadrons to be $\tau_0 = 8$~fm independent of $b$. Note that for
values $\tau_0 \sim 5$~fm or smaller, chemical equilibration is
practically complete for impact parameters smaller than 10~fm.

We now ask if such fluctuations can survive freezeout. If chemical
freezeout occurs before thermal freezeout, then baryons can suffer
elastic collisions with other hadrons after chemical fluctuations are
no longer possible.  Such scattering can restore $\cal C$ and $\cal V$
to thermal equilibrium values. To estimate the size of this effect, we
assume that those baryons and antibaryons that do not scatter
elastically contribute to fluctuations at the level of ${\cal V}_{\rm
  ch}$ and ${\cal C}_{\rm ch}$, while baryons that scatter once or
more are instantly thermalized. We then write \cite{rio},
\begin{equation}
{\cal V} \sim {\cal P}{\cal V}_{\rm ch} + (1-{\cal P}){\cal V}_{\rm th} 
\label{eq:fo1}
\end{equation}
where the survival probability for elastic scattering is
\begin{equation}
{\cal P} \sim {\rm e}^{-\int_{\tau_F}^\infty dt \;\sigma v_{\rm rel}
n_{\rm tot}(\vec{r}+\vec{v}t, t)},
\label{eq:fo2}
\end{equation}
for $n_{\rm tot}$ the hadron density, $\sigma$ the elastic
scattering cross section and $v_{\rm rel}$ the relative velocity. Pions 
dominate the late-time dynamics, so that $\sigma \approx
\sigma_{\pi N}\approx 20$~mb. To estimate $\cal P$,
we assume that after chemical freezeout the meson and baryon
distributions are described by spherically symmetric Gaussian
profiles for both position and velocity. Averaging over these
distributions, we find 
$0.8 < {\cal P} < 0.9$ for central collisions,
assuming $n_{\rm tot}(\tau_F)\approx
0.1$~fm$^{-3}$ and $1 < v_s\tau_F/R < 2$.
  
These calculations suggest that scattering after chemical freezeout
increases the chemical equilibrium variances and covariance by less
than $20\%$. Scattering at that level is not likely to wipe out the
large chemical fluctuations in figs.~1 and 3.  Furthermore, UrQMD
calculations suggest that chemical freezeout for baryons at RHIC may
occur much later than we have assumed \cite{urqmdHydro}.  In that
case, we expect rescattering to be completely negligible.

In summary, we have found that baryon density fluctuations lead to
striking correlations of protons and antiprotons if RHIC collisions
approach chemical equilibrium.  Measurements of the variance and
covariance of these species can reveal these correlations, even if
equilibration is incomplete and chemical freezeout precedes thermal
freezeout. We have not examined how the $\overline{p}$ and $p$
measurements are altered by resonance decays or hyperon contributions
(see \cite{JeonKoch} for a discussion of resonances in another
context). In addition, we have neglected mean field and coulomb
effects, which modify correlations over times $\gg \tau_F$.  Such
effects are important when the momentum dependence of
correlations is resolved in an HBT-like analysis, see
e.g.~\cite{pratt} and refs. therein. Our effect contributes
pre-freezeout correlations, often neglected in HBT discussions.

We are grateful to R. Bellwied, P. Fachini, A. Makhlin, R. Venugopalan
and N. Xu for discussions.  S.G. thanks the nuclear theory group at
Brookhaven National Laboratory for hospitality during part of this
work. This work is supported in part by the U.S. DOE grant
DE-FG02-92ER40713.

\end{narrowtext}
\end{document}